\documentclass{ws-procs9x6-cpt25}
\begin{document}

\title{From Explicit Diffeomorphism Breaking to Spontaneous Unimodularization}

\author{Yuri Bonder$^*$ and Johas D. Morales}
\address{Instituto de Ciencias Nucleares, Universidad Nacional Aut\'onoma de M\'exico\\
Circuito Escolar s/n, Ciudad Universitaria, 04510, CdMx, M\'exico\\
${}^*$bonder@nucleares.unam.mx}

\begin{abstract}
This contribution investigates modifications of General Relativity that allow or mimic energy nonconservation. We focus on Unimodular Gravity (UG), a theory that explicitly breaks diffeomorphism invariance, and show that it can lead to particle acceleration, opening the door to experimental constraints. We also argue that UG admits a well-posed initial value formulation, despite the presence of nondynamical structures. Finally, we introduce a simple model that spontaneously mimics UG, but in which a cosmological constant cannot arise as an integration constant.
\end{abstract}

\section{Introduction}

General Relativity (GR) is our most successful theory of gravity, validated across vast physical regimes.\cite{GRTests} Yet, GR faces profound challenges, especially its incompatibility with quantum mechanics. This points to a more fundamental theory predicting subtle deviations from GR, which can be sought by constructing new geometric gravitational models.

Energy conservation is a central principle in physics, but quantum mechanics doesn't strictly uphold it; e.g., a Hamiltonian's expectation value changes discontinuously during energy measurement due to state collapse. Moreover, GR lacks a well-defined gravitational energy notion, reinforcing the relaxation of this requirement for a fundamental gravity theory.

This raises the question: should theories bridging GR and quantum mechanics allow energy nonconservation? In GR, the conservation of the matter energy-momentum tensor, $T_{\mu\nu}$, is $\nabla^\mu T_{\mu\nu} = 0$, where $\nabla_\mu$ is the covariant derivative associated with the metric $g_{\mu\nu}$, and the metric (inverse metric) is used to lower (raise) spacetime indexes $\mu,\nu,\rho,\sigma,\ldots$. This contribution explores whether energy nonconservation can be incorporated into a geometric framework, with a focus on modified gravity theories that feature explicit or spontaneous breaking of invariance under diffeomorphisms.

\section{Unimodular gravity}\label{sec2}

Unimodular Gravity (UG) has been investigated by many authors.\cite{UG} The UG action, in $4$-spacetime dimensions, can be written as
\begin{equation}
S = \int d^4 x \, \frac{1}{2\kappa} \left[ \sqrt{-g} R + \lambda \left( \sqrt{-g} - F \right)\right] + S_M(g, \{\psi\}) ,
\end{equation}
where $\kappa = 8\pi G$, $G$ is Newton's constant (units where the speed of light is one are used), $R=g^{\mu\nu}R_{\mu\nu}$, with $R_{\mu\nu}$ the Ricci tensor associated with $g_{\mu\nu}$, and $g=\det(g_{\mu\nu})$. Moreover, \(\lambda\) is a Lagrange multiplier and $F$ is a \emph{nondynamical} scalar \emph{density}. Also, $S_M$ is the matter part of the action.

The equations of motion derived from this action include the ``unimodular constraint,'' $\sqrt{-g} = F$, the modified Einstein equation,
\begin{equation}\label{modified Einstein equation}
G_{\mu\nu} - \frac{1}{2} \lambda g_{\mu\nu} = \kappa T_{\mu\nu},
\end{equation}
where $G_{\mu\nu}= R_{\mu\nu}- (1/2)g_{\mu\nu}R$, and the matter field equations. Equation \eqref{modified Einstein equation} resembles the Einstein equation of GR, with $\lambda$ playing the role of a cosmological constant. However, there is no reason for $\lambda$ to be constant.

Equation \eqref{modified Einstein equation} can be manipulated by taking its trace, which yields
\begin{equation}\label{Tr modified Einstein equation}
 \lambda  =-\frac{1}{2}\left(R+\kappa T\right),
\end{equation}
where $T\equiv g^{\mu\nu}T_{\mu\nu}$. Inserting Eq.~\eqref{Tr modified Einstein equation} into Eq.~\eqref{modified Einstein equation} leads to
\begin{equation}\label{traceless Einstein}
 G_{\mu\nu} + \frac{1}{4} R g_{\mu\nu} = \kappa \left(T_{\mu\nu} - \frac{1}{4}g_{\mu\nu}T\right),
\end{equation}
which is the trace-free Einstein equation. This equation is interesting because, in a semiclassical approach, and under homogeneous and isotropic conditions, the vacuum expectation value of the energy-momentum tensor (derived from quantum fields) is proportional to the metric.\cite{Benito} Hence, its trace-free part vanishes, ensuring its absence from Eq.~\eqref{traceless Einstein}.

Moreover, taking the divergence of Eq.~\eqref{traceless Einstein} yields
\begin{equation}
 \nabla_\nu R  = \kappa \left(4 j_\nu - \nabla_\nu T \right),
\end{equation}
where the Bianchi identity is used and $j_\mu \equiv \nabla^\nu T_{\mu\nu}$. Thus, under the additional assumption that $j_\mu = 0$, it implies
\begin{equation}\label{trace of the div}
R  + \kappa T = 4\Lambda, 
\end{equation}
where $\Lambda$ is an arbitrary integration constant. When Eq.~\eqref{trace of the div} is substituted into Eq.~\eqref{traceless Einstein}, it becomes
\begin{equation}\label{Einstein equation with cosmological constant}
 G_{\mu\nu} + \Lambda g_{\mu\nu} = \kappa T_{\mu\nu},
\end{equation}
which is the conventional Einstein equation with a cosmological constant. Notably, the cosmological constant here arises as an integration constant, decoupled from matter field vacuum state, thereby offering UG as a solution to the cosmological constant problem.\cite{Weinberg} Also, when $j_\mu \neq 0$, the effective cosmological constant can be estimated from energy nonconservation mechanisms, yielding observationally consistent values.\cite{Daniel}

It can be verified that the presence of the nondynamical scalar density $F$ breaks diffeomorphism invariance, as is typically the case with nondynamical objects.\cite{Cristobal} The variation under an infinitesimal diffeomorphism, generated by the vector field $\xi^\mu$, of the gravitational action, $S_G=S-S_M$, yields
\begin{equation}
\delta S_G = \int d^4 x \, \frac{1}{2\kappa} \left[-2 \sqrt{-g}\left(G_{\mu\nu} - \frac{1}{2} \lambda g_{\mu\nu} \right)\nabla^{\mu}\xi^{\nu}+ \xi^\mu \nabla_\mu \lambda \left( \sqrt{-g} - F \right)\right].
\end{equation}
The first and last terms cancel due to the Bianchi identity (after integrating by parts) and the unimodular constraint, respectively. Therefore, for the action to be invariant, it is necessary that the diffeomorphisms be along vector fields subject to $\nabla_\mu \xi^\mu = 0$, which is consistent with the interpretation of $F d^4x$ as a preferred spacetime volume element. Importantly, any divergence-free vector field can be written in terms of an \emph{arbitrary} antisymmetric tensor field \(\alpha_{\mu\nu}\) as
\begin{equation}\label{div free}
\xi^\mu = \epsilon^{\mu\nu\rho\sigma} \nabla_\nu \alpha_{\rho\sigma},
\end{equation}
where \(\epsilon_{\mu\nu\rho\sigma}\) is the volume form, satisfying \(\nabla_\lambda \epsilon_{\mu\nu\rho\sigma} = 0\). 

Requiring the matter action to be invariant under diffeomorphisms generated by divergence-free vector fields leads to
\begin{equation}
\delta S_M = \int d^4 x \sqrt{-g} \, \alpha_{\sigma\lambda} \epsilon^{\nu\rho\sigma\lambda} \nabla_\rho \nabla^\mu T_{\mu\nu} = 0,
\end{equation}
where two integrations by parts are performed. Since this last equation is valid for all \(\alpha_{\mu\nu}\), it implies
\begin{equation}
0=\nabla_{[\nu} \nabla^\mu T_{\rho]\mu} = \nabla_{[\nu} j_{\rho]},
\end{equation}
where a pair of indexes enclosed in square brackets is antisymmetrized with a factor of $1/2$. A simple solution satisfying this condition is the usual conservation law: $j_\mu=0$. Yet, this theory allows for more general matter behaviors. In the absence of topological obstructions, as assumed here, the most general solution is
\begin{equation}
j_\mu = - \nabla_\mu \Phi,
\end{equation}
where \(\Phi\) is an arbitrary scalar called the energy-loss potential. Note that the action does not determine the energy-loss potential explicitly; it merely allows for the possibility of a nontrivial one. In UG, the potential can be introduced as additional input. An energy-loss potential is proposed and tested in the next subsection.

\subsection{Empirical Tests of Unimodular Gravity}

Tests involving particles and light have been extended to probe the energy loss allowed by UG.\cite{BonderHerreraRubiol} By generalizing Papapetrou's method,\cite{Papapetrou} the equation of motion for a pointlike test particle described by a nonconserved energy-momentum tensor, \(T_{\rm part}^{\mu\nu}\), is
\begin{equation}
u^\nu \nabla_\nu u^\mu = J^\nu \left( \delta^\mu_\nu + u^\mu u_\nu \right),
\end{equation}
where
\begin{equation}
J^\mu \equiv \frac{\lim \int_{\Sigma} d^3 x \, \sqrt{h} \, \nabla_\nu T_{\rm part}^{\mu\nu}}{\lim \int_{\Sigma} d^3 x \, \sqrt{h} \, T_{\rm part}^{tt}},
\end{equation}
and \(d^3 x \, \sqrt{h}\) is the volume element on a spacelike hypersurface \(\Sigma\) in Fermi coordinates associated with the particle's trajectory. The limits in the numerator and denominator arise because the method requires the particle to have finite extension during the derivation, with the point-particle limit taken only at the end.

Assuming spherical symmetry and that the energy-loss potential depends only on the radial coordinate, \(\Phi = \Phi(r)\), the system effectively reduces to a single degree of freedom. The motion is then governed by the effective energy equation
\begin{eqnarray}
E - \frac{1}{2} &=& \frac{1}{2} \dot{r}^2 + V_{\rm eff}, \\
V_{\rm eff} &=& \frac{1}{2 g_{rr}} \left( \frac{\ell^2}{r^2} - \kappa \right) + \Phi(r) - \frac{1}{2}.
\end{eqnarray}
Here, \(\ell\) is the conserved angular momentum associated with the Killing vector \(\partial_\phi\), while \(E\) is an integration constant. Note that there is no timelike Killing vector \(\partial_t\). Moreover, \(\kappa = -1, 0\) indicates whether the particle follows a timelike or null trajectory, respectively.

In addition, a Birkhoff-like theorem holds: if the traceless part of the Ricci tensor vanishes and spherical symmetry is imposed, the spacetime must be static.\cite{BonderHerreraRubiol} The corresponding line element is
\begin{equation}
ds^2 = -\left(1 - \frac{2M}{r} +  \tilde{\Lambda} r^2\right) dt^2 + \frac{dr^2}{1 - \frac{2M}{r} + \tilde{\Lambda} r^2} + r^2 d\Omega^2, 
\end{equation}
where \(M\) and \(\tilde{\Lambda}\) are integration constants, interpreted as the gravitational mass of the source and an effective cosmological constant, respectively. The latter is neglected in the remainder of this work.

An energy-loss potential of the form
\begin{equation}
\Phi(r) = \frac{b_1^{\rm t}}{r} + \frac{b_2^{\rm t}}{r^2} + \frac{b_3^{\rm t}}{r^3},
\end{equation}
is considered, where \(b_1^{\rm t}\), \(b_2^{\rm t}\), and \(b_3^{\rm t}\) are model parameters chosen to reproduce the structure of the extrema in \(V_{\rm eff}\). The deflection angle, \(\delta \phi\), for bound trajectories with semi-major axis \(a\) and eccentricity \(\epsilon\) is approximated by
\begin{equation}
\delta \phi = \frac{6 \pi M}{a (1 - \epsilon^2)} \left( 1 - \frac{b_2^{\rm t}}{3 M^2} \right) - \frac{6 \pi b_3^{\rm t}}{M a^2 (1 - \epsilon^2)^2}.
\end{equation}
Notably, \(\delta \phi\) is insensitive to \(b_1^{\rm t}\). For the specific case of Mercury, the prefactor \({6 \pi M}/{[a (1 - \epsilon^2)]}\) corresponds to the well-known perihelion precession of approximately 43 arcsec per century, where $M$ is the mass of the Sun.

Using PPN bounds,\cite{PPN} the observed deflection angle is constrained by
\begin{equation}
\delta \phi^{\rm PPN} = 43'' \left( 1 + \frac{2(\gamma - 1) - (\beta - 1)}{3} \right),
\end{equation}
where
\begin{eqnarray}
\gamma - 1 &=& (2.1 \pm 2.3) \times 10^{-5}, \\
\beta - 1 &=& (-4.1 \pm 7.8) \times 10^{-5}.
\end{eqnarray}
These constraints imply the bounds
\begin{equation}
-6.8 < (3.3 \times 10^{4}) \frac{b_2^{\rm t}}{M^2} + 1.1 \frac{b_3^{\rm t}}{M^3} < 1.4.
\end{equation}
Similar bounds hold for the case \(\kappa = 0\), which is relevant for light bending.\cite{BonderHerreraRubiol} This demonstrates that UG has observable empirical consequences. However, to establish UG as a viable theory, it must be predictive, an aspect explored in the next section.

\subsection{Initial value formulation}

Any viable physical theory must be predictive, or, in GR's argot, possess a well-posed initial value formulation. This requires identifying all constraints on the initial data and ensuring that the set of constraints are preserved under evolution. Furthermore, given initial data compatible with these constraints, the theory must yield a unique physical solution that depends continuously and causally on the initial data.\cite{wald1984general}

The well-posedness of the initial value formulation in UG was analyzed.\cite{HerreraBonder2024} It was shown that vacuum UG possesses a primary constraint, reminiscent of GR's momentum constraint. Requiring the preservation of this constraint under time evolution leads to a secondary constraint, analogous to the Hamiltonian constraint in GR with a cosmological constant. No further constraints are needed.

Moreover, using the BSSN formalism,\cite{BSSN} it was shown that even in the presence of a nontrivial energy-loss potential, the system remains hyperbolic. This, in turn, guarantees that the evolution is unique, continuous, and causal. Importantly, the unimodular condition, \(F = \sqrt{-g}\), can be enforced through an appropriate choice of the lapse function. Finally, sufficient conditions on the matter action to maintain this well-posedness were identified, which are similar to those required in GR.

Still, these results suggest that a well-posed initial value problem exists because the nondynamical object in UG behaves essentially as pure gauge. If, instead, the nondynamical object had true physical content, it could potentially obstruct the initial value problem, as there would be no mechanism ensuring that its future values remain compatible with the dynamical evolution. We turn to study spontaneous unimodularization.

\section{Spontaneous unimodularization}

Inspired by a question during the CPT'25 meeting, a theory replicating UG's features but spontaneously breaking diffeomorphism invariance is proposed, coining the term \emph{spontaneous unimodularization}. For that purpose, one can propose the action
\begin{equation}\label{teoriaoriginal}
S = \int d^4x \sqrt{-g}\left[\frac{1}{2\kappa}R-\frac{1}{2}\nabla^\mu\left(\frac{F}{\sqrt{-g}}\right)\nabla_\mu \left(\frac{F}{\sqrt{-g}}\right) - V \right]+S_M(g,\psi),
\end{equation}
where here $F>0$ is a \emph{dynamical} scalar density. Note that $F/\sqrt{-g}$ behaves like a scalar field and that the potential $V = V(F/\sqrt{-g})$ is chosen to drive $F$ towards $\sqrt{-g}$. The variation of Eq.~\eqref{teoriaoriginal} leads to
\begin{align}\label{1}
\kappa T_{\mu\nu} &= G_{\mu\nu} -\kappa \nabla_\mu \phi \nabla_\nu\phi +\kappa g_{\mu\nu}\left[\frac{1}{2}\nabla_\rho \phi \nabla^\rho\phi + V+ \phi (\Box \phi -  V' )\right],\\
\label{2}
\Box\phi &=V',
\end{align}
where the first equation is the Einstein equation and the second equation comes from the variation with respect to $F$. Here, we have abbreviated $\phi \equiv F/\sqrt{-  g}$, $\Box\phi = \nabla_\mu\nabla^\mu \phi$, and the prime represents the derivative with respect to the function's argument. Remarkably, on shell, the last two terms in Eq.~\eqref{1} cancel out and the dynamics is analogous to GR minimally coupled to a self-interacting Klein-Gordon field $\phi$.

The strategy is to substitute Eq.~\eqref{2} into Eq.~\eqref{1} and follow the same steps as in UG. Taking the trace of the resulting equation yields
\begin{equation}\label{trace}
\kappa T = -R +\kappa \nabla_\mu \phi \nabla^\mu\phi +4\kappa  V.
\end{equation}
Solving for \( V \) and substituting it back leads to
\begin{equation}\label{tracelessequation}
\kappa\left( T_{\mu\nu} - \frac{1}{4} g_{\mu\nu} T \right) = G_{\mu\nu} + \frac{1}{4} g_{\mu\nu} R - \kappa \left( \nabla_\mu \phi \nabla_\nu \phi - \frac{1}{4} g_{\mu\nu} \nabla_\rho \phi \nabla^\rho \phi \right),
\end{equation}
which is traceless and resembles Eq.~\eqref{traceless Einstein}, up to the scalar field \( \phi \) contribution. The divergence of Eq.~\eqref{tracelessequation} is
\begin{equation}\label{tracelessequation1}
- \frac{\kappa}{4} \nabla_\nu T= \frac{1}{4} \nabla_\nu R - \kappa \left[\nabla^\mu( \nabla_\mu \phi \nabla_\nu \phi) - \frac{1}{4} \nabla_\nu(\nabla_\rho \phi \nabla^\rho \phi) \right],
\end{equation}
where it is used that, since this model is diffeomorphism invariant, \( j_\mu \) has to vanish. Note that the expression inside the brackets on the right-hand side can be manipulated as follows:
\begin{eqnarray}
\nabla^\mu( \nabla_\mu \phi \nabla_\nu \phi) - \frac{1}{4} \nabla_\nu(\nabla_\rho \phi \nabla^\rho \phi) &=& \Box \phi \nabla_\nu \phi + \frac{1}{2} \nabla^\mu \phi \nabla_\mu \nabla_\nu \phi \nonumber\\
&=& V' \nabla_\nu \phi + \frac{1}{2} \nabla^\mu \phi \nabla_\mu \nabla_\nu \phi \nonumber\\
&=& \nabla_\nu\left( V + \frac{1}{4} \nabla^\mu \phi \nabla_\mu \phi \right),
\end{eqnarray}
where we have used Eq.~\eqref{2} and properties of the covariant derivative. Consequently, Eq.~\eqref{tracelessequation1} can be written as
\begin{equation}\label{tracelessequation2}
0 = \nabla_\nu\left( R + \kappa T - 4\kappa V - \kappa \nabla^\mu \phi \nabla_\mu \phi \right),
\end{equation}
whose solution, similarly to UG, is
\begin{equation}\label{tracelessequation3}
4\Lambda = R + \kappa T - 4\kappa V - \kappa \nabla^\mu \phi \nabla_\mu \phi.
\end{equation}

However, in stark contrast to UG, the separation between the trace and traceless parts of the Einstein equation is artificial in this case, and the trace equation is nontrivial. As a result, the integration constant $\Lambda$ is not arbitrary but can be explicitly determined. In fact, it follows directly from Eq.~\eqref{trace} that $\Lambda = 0$, implying that the mechanism of spontaneous unimodularization fails to generate a cosmological constant from an integration constant.

Still, it is possible to follow what is done in UG and introduce Eq.~\eqref{tracelessequation3} into Eq.~\eqref{tracelessequation}, which produces
\begin{equation}\label{tracelessequation4}
\kappa\left( T_{\mu\nu} - \frac{1}{2} g_{\mu\nu} T \right) = R_{\mu\nu} - \kappa \nabla_\mu \phi \nabla_\nu \phi  - g_{\mu\nu}\kappa V,
\end{equation}
and whose trace reversed version is
\begin{equation}\label{tracelessequation5}
\kappa T_{\mu\nu} = G_{\mu\nu} +\kappa g_{\mu\nu}  V - \kappa \nabla_\mu \phi \nabla_\nu \phi +\frac{1}{2}g_{\mu\nu} \kappa \nabla^\mu \phi \nabla_\mu \phi.
\end{equation}
This equation is the Einstein equation for the case where there is matter and an additional self-interacting field $\phi$.

When $V$ has a ``deep'' minimum and the field ``lies'' at the minimum, it acquires a value $\phi_0$ such that $ \nabla_\mu \phi_0 $. Therefore, Eq.~\eqref{tracelessequation5} becomes
\begin{equation}\label{tracelessequation5}
\kappa T_{\mu\nu} = G_{\mu\nu}+ g_{\mu\nu}\Lambda_{\rm eff},
\end{equation}
where $\Lambda_{\rm eff}=\kappa V(\phi_0$ acts as an effective cosmological constant. Notice that it is possible to use any potential with a minimum. However, the field may fluctuate around the minimum, producing interesting physics that are similar to what has been discussed extensively in the context of inflation.\cite{scalarinflation}

\section{Conclusions}

The introduction of nondynamical objects generally leads to unconventional physics, including the breaking of diffeomorphism invariance and energy nonconservation. In UG, this symmetry breaking opens the door for particular particle dynamics, which allows one to place bounds on possible mechanisms of energy loss. On the other hand, studying the initial value formulation of theories containing nondynamical objects is of limited value, since these objects have fixed ``values'' that cannot be predicted dynamically. However, UG represents a notable exception. In this theory, the nondynamical scalar density can be effectively absorbed into the lapse function. This feature provides a natural explanation for why UG constitutes such a mild deviation from GR, maintaining a well-posed initial value problem.

To explore whether UG phenomenology can be reproduced without invoking energy nonconservation, we considered a minimally coupled, self-interacting Klein-Gordon field $\phi$, defined as the ratio of a dynamical scalar density $F$ to $\sqrt{-g}$, and subject to a potential that induces spontaneous unimodularization. Upon taking the divergence of the traceless part of Einstein equation, we found that \emph{no} integration constant emerges that could be reinstated into Einstein’s equation. Nevertheless, when the field settles at the minimum of its potential, its value contributes to the effective cosmological constant; an effect reminiscent of the mechanism responsible for scalar-field-driven inflation.

\section*{Acknowledgements}
We thank J.E. Herrera and A.M. Rubiol for their collaboration in Sec.~\ref{sec2}. This work was supported by UNAM DGAPA-PAPIIT grant IN101724.

\end{document}